\documentclass[11pt,a4paper]{article}
\usepackage{amsmath}
\usepackage{dsfont}
\usepackage{bm}
\usepackage{esint}
\usepackage{subcaption}
\usepackage{accents}
\usepackage{mathrsfs}
\usepackage{amsbsy}
\usepackage{subfiles}

\usepackage{a4wide,graphicx,times,psfrag,wrapfig,sidecap}
\usepackage{cite}
\usepackage[colorlinks=true,linkcolor=black, citecolor=black,
urlcolor=black]{hyperref}
\numberwithin{equation}{section}
\makeatletter \let\old@startsection=\@startsection
\renewcommand{\@startsection}[6]
{\old@startsection{#1}{#2}{#3}{#4}{#5}{#6\mathversion{bold}}}
\makeatother
\def\Res{ \rm{Res}}

\def\<{\langle}
\def\>{\rangle}
\def\sign{{\rm sign}}
\def\tr{{\rm   tr} }
\def\Im{{\rm Im}}
\def\Re{{\rm Re}}
\newcommand\encadremath[1]{\vbox{\hrule\hbox{\vrule\kern8pt
\vbox{\kern8pt \hbox{$\displaystyle #1$}\kern8pt}
\kern8pt\vrule}\hrule}} \def\enca#1{\vbox{\hrule\hbox{
\vrule\kern8pt\vbox{\kern8pt \hbox{$\displaystyle #1$} \kern8pt}
\kern8pt\vrule}\hrule}}
  \usepackage{bm}

\def\Xint#1{\mathchoice
   {\XXint\displaystyle\textstyle{#1}}%
   {\XXint\textstyle\scriptstyle{#1}}%
   {\XXint\scriptstyle\scriptscriptstyle{#1}}%
   {\XXint\scriptscriptstyle\scriptscriptstyle{#1}}%
   \!\int}
\def\XXint#1#2#3{{\setbox0=\hbox{$#1{#2#3}{\int}$}
     \vcenter{\hbox{$#2#3$}}\kern-.5\wd0}}

\def\dashint{\Xint-}

\usepackage{amssymb}

\begin{document}

\thispagestyle{empty}
\vspace{1cm}
\setcounter{footnote}{0}
\begin{center}
{\Large\bf Exact Matrix Elements of the Field Operator in the Thermodynamic Limit of the Lieb-Liniger Model }
\end{center}
\begin{center}
Eldad Bettelheim \\[7mm]

{\it Racah Inst.  of Physics, \\Edmund J. Safra Campus,
Hebrew University of Jerusalem,\\ Jerusalem, Israel 91904 \\[5mm]}
{\bf Abstract}
\end{center}
\noindent{ We study a matrix element of the field operator in the  Lieb-Liniger model using the Bethe ansatz technique coupled with a functional approach to compute Slavnov determinants. We obtain the matrix element exactly in the thermodynamic limit for any coupling constant $c$, and compare our results to known semiclassics at the limit $c\to0$  }

\section{Introduction}
The Lieb-Liniger model of one-dimensional  Bosons with delta function interaction, is important both from a perspective geared to understanding real quantum interacting systems and from a mathematical physics perspective, as the model is exactly solvable through the Bethe ansatz. The semiclassical limit of the model is also of  importance, since this limit describes  classical integrable systems of the non-linear Schr\"{o}dinger type (also known as the Gross-Pitaevskii model) which find applications in several branches of physics. As the problem is "solvable", much can be said exactly about the behavior of systems described by these quantum and classical models. In particular, the quantum problem enjoys a full solution for the thermodynamic potentials making use of the Bethe ansatz, but as is usually the case, the computation of matrix elements, expectation values and quantum overlaps through the Bethe ansatz is more difficult. In this paper we give an expression for the matrix elements of the field operator in the Lieb-Liniger model, a quantity which we believe is key in understanding the model beyond equilibrium. The result we present here may be extended in several direction offering perhaps a much fuller understanding of the Lieb-Linger model in and out of equilibrium.

To compute the matrix elements of the field operator in the Lieb-Linger  model we first use a representation of this matrix elements as a sum over certain overlaps of Bethe states \cite{Korepin:Bogoliubov:Izergin:Quantum:Inverse:Scattergin:p264,Calabrese:Lieb:Liniger:Form:Factors} (one an eigenstate of the Hamiltonian and the other one not), which are then expressed using Slavnov determinants \cite{Slavnov,Calabrese:Lieb:Liniger:Form:Factors}.  

As mentioned above we use the functional approach \cite{Kostov:Bettelheim:SemiClassical:XXX,Gorohovsky:Bettelheim:A:Derivation,Bettelheim:Kondo} to study these Slavnov matrix elements. In this approach, the Slavnov matrix, the determinant of which must be computed, is written in terms of an operator acting on functions with certain analytical properties in the complex plane. This operator is then inverted by solving an operator equation. We achieve this inversion here by making use of the Steepest descent method. Indeed, the operator which we deal with is an integral operator with a Kernel that contains an exponent of the form $e^{N \imath \varphi(z)}$, where $\varphi(z)$ is a function which has a finite   thermodynamic limit, and $N$\ is the number of particles. Thus the integral operator can be deal with within the saddle point method.

We should mention that similar computations but for expectation values and matrix elements were performed in Refrs  \cite{Caux:Calabrese:Slavnov:1P:Dynamical:in:LiebLinger,Calabrese:Piroli:Alvise:Exact:Correlations:Lieb:Liniger,De:Nardis:Panfil:Form:Factors:Lieb:Liniger,Kormos:Mussardo:Trombettoni:Expectations:Lieb:Liniger,Kormos:Mussardo:Trombettoni:Lieb:Liniger:NonRelativistic,Koubek:Mussardo:Operator:Content:Sinh:Gordon}, our method is more general in that it refers to the field operator in the model from which the density operator can be formed. We also believe that the method shown here, although applied in this instance only to the Lieb-Liniger model, may be generalized to treat important problems in other Bethe-solvable models.   

  After solving the inversion problem of the Slavnov matrix, we compute the determinant by the usual means, roughly speaking making of the very general identity $\det O(\alpha)=\exp\int \tr O^{-1}(\alpha) dO(\alpha). $ Upon obtaining the result, we take the small $c$ limit, in order to be able to compare our results  to known semiclassics. In fact such a program has been undertaken in  Refrs. \cite{Gorohovsky:Bettelheim:A:Derivation,Kostov:Bettelheim:SemiClassical:XXX}, however in those papers, $c$ was taken to be of order $1/N$ or smaller, which is not the approach taken here. The advantage of our approach is that we can use our techniques to find the matrix elements for finite $c$, whereas Refrs. \cite{Gorohovsky:Bettelheim:A:Derivation,Kostov:Bettelheim:SemiClassical:XXX} are limited to small $c$, and thus {\it restricted} to the semiclassical limit. In our approach however we do not reproduce the full semiclassical result, due to the fact that, in order to use the steepest descent method, we cannot let $c$ be of order $1/N$. Nevertheless, except a certain phase space element,  the other essential features of semiclassics are reproduced in our result. 

We thus, after a formal introduction to the Lieb-Linger model, to fix notations, we begin with the computation, and discuss further the results and their interpretation in the conclusions.    

\section{The Lieb-Liniger Model and Associated  Slavnov Determinant }

The Lieb-Liniger model on a lattice may be defined using the following $L$-matrix:
\begin{align}
L(n,\lambda)=\mathds{1}+dx\left(\begin{array}{cc}
-\frac{\imath\lambda  }{2} & \imath \sqrt{c}\psi^\dagger_n \\ 
\imath \sqrt{c}\psi_n & \frac{\imath\lambda  }{2}
\end{array} \right)
\end{align}
It is then possible to show the following:
\begin{align}
R_{jj'}^{ii'}(\lambda,\mu)L_{i'k}(n,\lambda)L_{j'l}(n,\mu)=L_{ik'}(n,\mu)L_{jl'}(n,\lambda)R_{l'l}^{k'k}(\lambda,\mu),
\end{align}
where 
\begin{align}
R(\lambda,\mu)=\begin{array}{c} {\scriptstyle \,^1_1}\\{\scriptstyle\,^1_2}\\{\scriptstyle\,^2_1}\\{\scriptstyle \,^2_2}\end{array}
\overset{\,^1_1\quad\quad\quad^1_2\quad\quad^2_1\quad\quad\quad^2_2\,}{\left(\begin{array}{cccc}
1+\frac{\imath c}{\mu-\lambda} & 0 &  0 & 0 \\
0 &\frac{\imath c}{\mu-\lambda} &1 & 0 \\
0&1 &\frac{\imath c}{\mu-\lambda} &0 \\
0& 0 & 0 & 1+\frac{\imath c}{\mu-\lambda}
\end{array}\right)}
\end{align}
namely what is meant here that writing a tensor as a matrix we have always:
\begin{align}
A=\left(\begin{array}{cccc}
A^{11}_{11} & A^{11}_{12} & A^{12}_{11} & A^{12}_{12} \\
A^{11}_{21} & A^{11}_{22} & A^{12}_{21} & A^{12}_{22} \\
A^{21}_{11} & A^{21}_{12} & A^{22}_{11} & A^{22}_{12} \\
A^{21}_{21} & A^{21}_{22} & A^{22}_{21} & A^{22}_{22}  \end{array}\right)
\end{align}

The Bethe equations for this model read:
\begin{align}
e^{\imath L\theta_j}\prod_{k=1}^N\frac{\theta_j-\theta_k-\imath c}{\theta_j-\theta_k+\imath c}=-1\label{Bethe}
\end{align}

We wish to compute within this model the following object:
\begin{align}
\frac{\<{\rm out}|\psi|{\rm in}\>}{\sqrt{\<{\rm in}|{\rm in}\>\<{\rm out}|{\rm out}\>}},
\end{align}
where $|{\rm in}\>$ and $|{\rm out}\>$ are two Bethe states (on-shell). This expression may be brought into the framework of the inverse scattering method with the help of  the following identity  \cite{Calabrese:Lieb:Liniger:Form:Factors,Korepin:Bogoliubov:Izergin:Quantum:Inverse:Scattergin:p264}:
\begin{align}
\psi \prod_{i=1}^N B(\theta_i)|0\>=-\imath \sqrt{c} \sum _ke^{-\frac{\imath L\theta_k}{2}}   \prod _{j(\neq k)} \frac{\theta_k-\theta_j+\imath c}{\theta_k-\theta_j}B(\theta_j)|0\>.\label{removals}
\end{align}
To compute the ensuing matrix elements we use the Kostov-Matsuo formula:
\begin{align}
\<\tilde{\bm{\theta}}|\bm{\theta}\>=e^{\frac{\imath L}{2}\sum_j \theta_j-\tilde \theta_j}\det( \mathds{ 1} -  K)   
\end{align}
In this paper  we will use  a new  determinant  representation,
where the $N\times N$ matrix $K$ is  defined as  
\begin{align} 
&K_{ij} =E_i\frac{1  }{u_{i} -u_j+ \imath c }\label{Kdef}  \\
&E_i \equiv e^{-\imath Lu_i} 
  {\prod_{k}(u_{i} -u_k+ \imath c)    \over \prod_{k\neq i}(u_{i}-u_k)}
\end{align}
and $\bm u\equiv\{u_i\}_{i=1}^{2N}$ is the union of $\bm\theta$ and $\tilde {\bm \theta}$, which are assumed to be disjoint: $\bm\theta\cap\tilde{\bm\theta}=\emptyset$, $\bm u=\bm\theta\cup\tilde{\bm \theta}$. 

First we define as the following expression the object we wish to compute:
\begin{align}
&\frac{\<\bm{\theta}^{\rm out}|\psi(0)|\bm \theta ^{\rm in}\>}{\sqrt{\<\bm{\theta}^{\rm out}|\bm{\theta}^{\rm out}\>\<\bm \theta ^{\rm in}|\bm \theta ^{\rm in}\>}}\label{tocompute}=\\&
\nonumber=\frac{-\imath \sqrt{c}e^{\frac{\imath L}{2}\sum_j \theta^{\rm out}_j-\theta^{\rm in}_j}\sum_{q_2}\det( \mathds{ 1} -  K_{\bm{\theta}^{\rm out}\cup\bm \theta ^{\rm in}\setminus\theta^{\rm in}_{q_2}})    \prod _{j\neq q_2} \frac{\theta^{\rm in}_{q_2}-\theta^{\rm in}_j+\imath c}{\theta^{\rm in}_{q_2}-\theta^{\rm in}_j}}{\sqrt{ \det( \mathds{ 1} -  K_{\bm{\theta}^{\rm out}\cup\bm{\theta}^{\rm out+\varepsilon}})\det( \mathds{ 1} -  K_{\bm{\theta}^{\rm in}\cup\bm{\theta}^{\rm in+\varepsilon}})}} 
\end{align}   
where $\bm \theta^{\rm out}$ has $N$ elements and $\bm \theta^{\rm in}$ has $N+1$  elements. The computation of the objects appearing in Eq. (\ref{tocompute}) are easy except the determinants. We will have a result for these determinants in Eq. (\ref{MainResultGeneral}), which will enable the computation of the matrix element in Eq. (\ref{tocompute}) for any $c$ of order $1$ namely for $c$\ that does not scale in any way with $N$\ at the $N\to\infty$ limit.

\section{Functional Approach to the Slavnov Determinant}

The matrix $K$\ can be given an operator form by the following trick. We encode any vector $\bm{v}$ in $\mathds{C}^{2N}$ by a function $[\bm v](\lambda)$ in the following way:
\begin{align}
[\bm v](\lambda)=\sum\frac{v_i}{\lambda-u_i}.
\end{align}
This simple mapping allows us to write $K$ in eq. (\ref{Kdef}) as:
\begin{align}
[K\bm v](\lambda)=\oint  \frac{e^{N\imath\varphi(\lambda')}v(\lambda'+\imath c)}{\lambda-\lambda'} \frac{d\lambda'}{2\pi\imath},
\end{align}
the contour integral to be taken around any contour surrounding the $u_i$'s but excluding the points $\xi_l-\frac{2\imath }{dx}$. The function $\varphi$ featuring in this formula is given by:
\begin{align}
e^{N\imath\varphi(\lambda)}=   e^{-\imath L\lambda} \prod_l
  {\lambda-u_l+ \imath c    \over \lambda-u_l}.\label{varphidef}
\end{align} 
If we denote by $\bm e^{(j)}$ the vector with elements $e^{(j)}_i=\delta_{ij}$ then the function \begin{align}R(\lambda,u_j)\equiv[(1-K)^{-1}\bm{e}^{(j)}](\lambda)\label{Rdef}\end{align} solves:
\begin{align}
R(\lambda,w)-\oint  \frac{e^{N\imath\varphi(\lambda')}R(\lambda'+\imath c,w)}{\lambda-\lambda'} \frac{d\lambda'}{2\pi\imath}=\frac{1}{\lambda-w},\label{ToSolve}
\end{align}
where $w$ should be set to $u_j$. The reason we write $w$ instead of $u_j$ is that this equation may be analytically continued\cite{Kostov:Bettelheim:SemiClassical:XXX} to any $w$ in a neighborhood of the $u_i$  and so we may wish to solve Eq. (\ref{ToSolve}) treating $R$ as an analytic function of  $w$  in a neighborhood of the $u_i$'s.

Let us pause to give some general properties of the function $\varphi$. First,

\begin{align}
\varphi'(z)=-\rho_0^{-1} +2\imath \int_{}^{} \left(\frac{1}{z-z'}-\frac{1}{z-z'+\imath c}  \right)\rho(z'){d z'}, \label{varphiprimefull} 
\end{align}
where $\rho$ is the occupation of mode number on the real axis. It is doubled here since both in- and out- states contribute in the symmetric representation of the overlaps used here. Comparison  with the Bethe equation, Eq. (\ref{Bethe}),  gives the first of the following equations while the second is derived from Eq. (\ref{varphiprimefull})  directly for $x$ real and $z,$ in general, complex:
\begin{align} 
&\Re(\varphi'(x\pm\imath0^+))=2\pi \rho(x)(-1\pm1)\label{phipmsaysPis0}.\\
&\Im(\varphi'(x\pm\imath 0^+))=\Im[\varphi'(x)]^A\label{phipmImsmooth}\\
&\varphi'(z-\imath c)=\overline{\varphi}'(z)\label{phiprmsaysImPnot0}
\end{align}
furthermore the imaginary part of $\varphi'$ is smooth across the real axis and across the line $\mathds{R}-\imath c$. Here the superscript $A$\ denotes taking the average value of the function from both sides of the real axis: $f^A(x)\equiv \frac{f(x+\imath0^+)+f(x-\imath0^+)}{2} $.

The trace is easily computed in terms of an object we term $R_y,$ which is derived from  $R$\ defined in Eq. (\ref{Rdef}), by replacing $K$ by $e^{-Ny}K.$ Namely, we take: \begin{align}R_y(\lambda,u_j)\equiv[(1-e^{-Ny}K)^{-1}\bm{e}^{(j)}](\lambda).\label{Rydef}\end{align} This allows us to write\begin{align}
&\tr (\mathds{1}-e^{-Ny}K)^{-1}=\sum_j \underset{\lambda\to u_j}{\Res} R_y(\lambda,u_j)=\oint\oint\frac{R_y(\lambda,w)}{\lambda-w}\frac{dz}{2\pi\imath}\frac{dw}{2\pi\imath} ,\label{trasint}
\end{align}
where the $w$ integral surrounds the $\lambda$ integral, which in turn surrounds the Bethe roots.
Now we can use (\ref{trasint}) to obtain:\begin{align}
&N+\log\det(\mathds{1}-K)= N\int_0^{\infty} \tr (\mathds{1}-e^{-Ny}K)^{-1}dy\label{actualmaineq}=\\&\nonumber= N\int_0^{\infty}\oint\oint\frac{R_y(z,w)}{z-w}\frac{dz}{2\pi\imath}\frac{dw}{2\pi\imath} dy.
\end{align}

Thus, in order to find the matrix elements, it remains to  compute $R_y(z,w)$ in the thermodynamic limit, which we shall do in the next sub-section.We shall be interested in solving the problem in Fourier space, which leads us first to consider the Legendre transform, $\tilde\varphi,$ of $\varphi $:
\begin{align}
\tilde \varphi(P(z))=\varphi(z)-P(z)z-\frac{\log P'(z)}{2N\imath },\quad\mbox{for}\quad P(z)\equiv\varphi'(z).
\end{align}         
Of course the Legendre transform is the exponent in the saddle point approximation of the Fourier transform of $\varphi$, if $P$\ is identified with $p/N$, for $p$ the Fourier variable:
\begin{align}
\oint e^{N\imath\varphi(z)-\imath  NPz }dz\simeq e^{  N \imath \tilde\varphi(P)}
\end{align}
similarly we define $\tilde {S}$ by the following: 
\begin{align}
\int R(z\pm\imath0^+,w)e^{-\imath NP x }dx\simeq e^{  N \imath \tilde S_\pm(P,w)}.
\end{align}
The equation for $\tilde{S}$ is then given by:
\begin{align}
e^{\imath N \tilde{S}_\tau (P,w)} -\int _0^\infty N e^{\imath N [\tilde{S}_+ (Q,w)+\imath cQ+\tilde{\varphi}(P-Q)]}dQ  =\delta_{\sigma,\tau}\sigma e^{-\imath N P w}\label{PIntegralEquation}
\end{align}
 for $\Im(w)>0,$ where $\sigma=\sign(P)$.

\subsection{Solution of the Integral Equations}

We take  equation  (\ref{PIntegralEquation}) and solve it by postulating the following solution:
\begin{align}
&e^{\imath N \tilde{S}_+ (P,w)} =\frac{\theta(P>0)e^{-\imath N P w}}{1-e^{\imath N(\varphi_+(w)+\imath c P)}} , \quad \Im(w)>0,\label{quasifullsolution}
\end{align}
where $\varphi_\pm$ are the analytic continuation of $\varphi$ from above or below the real axis, respectively. $\varphi$ has a jump discontinuity at the support of the Bethe roots, so that we have two different analytical continuations through the cut depending on whether one comes from above or below the real axis. In the end we shall take $w$ to be real, so that in the final equations for the matrix elements, $\varphi_\pm(w), $ will simply denote $\varphi(w\pm\imath0^+), $ respectively. 

Actually what we meant by Eq. (\ref{quasifullsolution}) is that there are two asymptotes:
\begin{align}
e^{\imath N \tilde{S}_+ (P,w)} =\begin{cases}-e^{-\imath N(\varphi_+(w)+(w+\imath c) P)} & 0<P<-\frac{1}{c}\Im(\varphi_+(w)) \\
e^{-\imath N P w} & 0<-\frac{1}{c}\Im(\varphi_+(w))<P \\
\end{cases}.\label{thesolution}
\end{align} 
Performing the inverse Fourier transform on this object one obtains the following form for ${R}(z,w)$:
\begin{align}
R(z,w)=\frac{e^{\frac{\imath N}{c}(z-w)\Im\varphi_+(w)}}{z-w}.\label{RfromSplus}
\end{align}
Here only $P$ values above $-\frac{\Im(\varphi_+(w))}{c}$ were taken into account (the second line on the right hand side of Eq. (\ref{thesolution})), since  values of $P$ lower than $-\frac{\Im(\varphi_+(w))}{c}$ add an exponentially small contribution with respect the former.

Let us we show that the solution given in Eq. (\ref{thesolution}) to Eq. (\ref{PIntegralEquation}) is valid. Indeed, plugging Eq. (\ref{thesolution}) into the integral on the left hand side of Eq. (\ref{PIntegralEquation}), and computing  the integral for $P$ positive, one obtains:
\begin{align}
& \int _0^\infty  e^{\imath N [\tilde{S}_+ (Q,w)+\imath cQ+\tilde{\varphi}(P-Q)]}
dQ\label{twointegrals}=\\&=e^{-\imath NPw }\left[e^{-\imath N  \varphi_+(w) }\int _{P}^{P'}  e^{\imath N [ Qw+\tilde{\varphi}(Q)]}dQ+e^{- NPc }\int _{P'}^{-\infty} e^{\imath N [ Q(w-\imath c)+\tilde{\varphi}(Q)]}dQ\right],\nonumber
\end{align}
where \begin{align}\Re(P')=P+\frac{\Im(\varphi_+(w))}{c}\label{RePp}.\end{align} A change of integration variable $Q\to P-Q $ is performed on the second line . We define now:
\begin{align}
H(z_1;z_2)=\varphi(z_1)- \varphi'(z_1)(z_1-z_2)\label{Hwdefinition}. 
\end{align}
This function has the property that the saddle point with respect to $z_1$ is achieved at the point $z_2$ and the value of the saddle point is $\varphi(z_2)$:
\begin{align}
  H(z_2;z_2)=\varphi(z_2),\quad H'(z_2;z_2)=0,\quad H''(z_2;z_2)=-\varphi''(z_2). 
\end{align}
By changing the integration variable from $Q$ to $z$ where $\varphi'(z)=Q$ one obtains:
\begin{align}
& \int _0^\infty  e^{\imath N [\tilde{S}_+ (Q,w)+\imath cQ+\tilde{\varphi}(P-Q)]}
dQ=\\&=e^{-\imath NPw }\left[e^{-\imath N  \varphi_+(w) }\int _{z(P)}^{z(P')}  e^{\imath N H(z,w)}\frac{dz}{z'} +e^{- NPc }\int _{z(P')}^{z(-\infty)} e^{\imath N H(z,w-\imath c)}\frac{dz}{z'}\right],\nonumber
\end{align}
where $z'$\ denotes the following $z$-dependent function  $z'\equiv \left.\frac{dz(P)}{dP} \right|_{P=P(z)}$ . 

\begin{figure}[h!!!]
\begin{center}
\includegraphics[width=8cm]{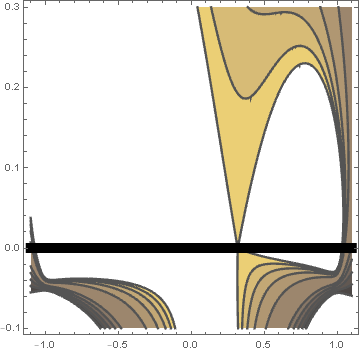}
\caption{
We set $w=0.2$, $\rho_0=1$, $c=0.1$. Black heavy line denote the branch cut.  Level line of $-\Im(H(z,w))$ are drawn the $z$ plane, for $w=0.315$. The level line with the value $-\Im H(w,w)=0.9$ touches down on the branch cut. All other level lines have smaller values and the shading is drawn such that brighter colors denote a higher value of $-\Im (H(z,w))$.  \label{Map}} \end{center}
\end{figure}

 Note that when switching from the first integral to the second integral, namely  when the following inequality  $\Re \,P(z)=\Re \varphi'(z)<\Re (P')$ sets hold, we get: 
\begin{align}
&\Re \left(\imath H(z,w-\imath c) -cP-\imath H(z,w)+\imath\varphi_+(w)\right)= c\Re \varphi_+'(z)-cP-\Im(\varphi_+(w)\nonumber)\\&<c\Re( P')-cP-\Im(\varphi_+(w))=0,\label{secondintegralirrelevant}
\end{align} 
the last equality being a consequence of Eq. (\ref{RePp}). Thus switching to the  second integral at $z(P')$ provides a smaller contribution than the result that would be achieved if one continues to integrate over the first integrand. This means that saddle points  reached within the second integral are exponentially suppressed. Since the saddle points in the first integrals correspond to $\varphi_+$ or  $\varphi_-$, respectively, the contributions related to $\varphi_-$ may be ignored. Fig. \ref{Map} shows the landscape on which the steepest descent curves are to be drawn.

The first integral  in Eq. (\ref{twointegrals}) is just the inverse Fourier transform of $\tilde{\varphi}$, so we obtain the following up to exponentially small terms:

\begin{align}
& \int _{z(P)}^{z(P')}  e^{\imath N H(z,w)}\frac{dz}{z'} =e^{\imath N  \varphi_+(w)}.
\end{align}
The reason that $\varphi_+(w)$ appears on the right hand side and not $\varphi_-(w)$ is that the saddle point reached within this integral corresponds to the analytical continuation of $H(z,w)=\varphi(z)- \varphi'(z)(z-w)$ from the upper half plane. The saddle point reached from analytically continuing $H(z,w)$ from the lower half plane, is not reached within this integral since the relevant momentum $P(w)=\varphi_-'(w)$ is too negative. Indeed, we shall show in the next paragraph that:
 \begin{align}
 \varphi'_-(w)<P' \label{minusinequality}.
 \end{align}
 
To show inequation (\ref{minusinequality}), we combine the definition of $\varphi$ with the Bethe equations, Eqs. (\ref{varphidef}, \ref{Bethe}), repsectively, to obtain:
\begin{align}
\Re(\varphi(\theta_i-\imath 0^+))=-\frac{2\pi}{N} (n_i+1)
\end{align}
where $n_i$ is the mode number of the Bethe root $\theta_i$. Taking the derivative of this equation and remembering that each mode number is represented twice, namely once for the in-state, and once more for the out-state, one obtains:
\begin{align}
\Re \varphi'(x-\imath0^+)=-4\pi\rho(x),\label{realphiprimeisrho}
\end{align} 
where $\rho(x)$ is the density of occupied mode numbers on the real axis. Now, as will be shown later in Eq. (\ref{phipmsaysPis0}), in the small $c$ limit one has, 
\begin{align}
&\frac{1}{c}\Im\varphi(x+\imath0^+)=-2\pi\rho(x),\label{Imphiisrho}
\end{align}
so that by combining Eqs. (\ref{realphiprimeisrho}, \ref{Imphiisrho}) and Eq. (\ref{RePp}) one obtains the inequality, Eq. (\ref{minusinequality}), given that $P$\ in Eq. (\ref{minusinequality}) is positive and in the small $c$ limit. For larger $c$ one may ascertain that the inequality holds for specific choices of $c$. 

We have given above, namely in Eqs. (\ref{thesolution}), the solution for $\tilde{S}_+,$   which leaves us to compute $\tilde{S}_-.$ In order to compute the latter, one may return to equation (\ref{PIntegralEquation}) and set $P$\ negative, which gives the following equation for $e^{\imath N\tilde{S}_-}$ \begin{align}
&e^{\imath N \tilde{S}_- (P,w)} =\int _0^\infty NdQ e^{\imath N [\tilde{S}_+ (Q,w)+\imath cQ+\tilde{\varphi}(P-Q)]}- e^{-\imath N P w}\end{align}
This can be further computed as follows:
\begin{align}
&e^{\imath N \tilde{S}_- (P,w)}\label{FindvSm}=\\&=e^{-\imath NPw }\left[e^{-\imath N  \varphi_+(w) } \int _{z(P)}^{z(P')}  e^{\imath N H(z,w)}\frac{dz}{z'}+e^{- NPc }\int _{P'}^{z(-\infty)} e^{\imath N H(z,w-\imath c)}\frac{dz}{z'}-1\right]\nonumber=\\
&=\frac{e^{\imath N(\tilde \varphi(P)
 -  \varphi_+(w)) }}{z(P)-w}  -e^{-\imath NPw }.\nonumber
\end{align}
The second integral on the second line is irrelevant due again to the inequality given in Eq. (\ref{secondintegralirrelevant}).
The first integral receives contributions from the limits of the integration domain, but no saddle point is reached. This edge contribution gives rise to the first term in the last row of Eq. (\ref{FindvSm}).

Taking the Fourier transform of Eq. (\ref{FindvSm}) one  obtains:
\begin{align}
R(z-\imath0^+,w)=\frac{e^{\imath N(\varphi (z)
 -  \varphi_+(w)) }-1}{z-w}.\label{RminFinal}
\end{align}
This result is of course also  obtained  again by the saddle point method. The saddle of integrating $e^{\imath N\tilde \varphi(P) +\imath Pz}$ is located of course at $z$. Depending on whether $z$ is in the upper half plane or the lower half plane, one obtains either $\varphi_+(z)$ or $\varphi_-(z)$ as a result, such that in Eq. (\ref{RminFinal}) one may simply write $\varphi(z)$. 

\subsection{Computing the trace}
We can now use Eq. (\ref{actualmaineq}) to compute the Slavnov determinant. The object $R_y(z,w)$ is simply $R(z,w)$ after one takes $\varphi(z)\to \varphi (z)+\imath  y,$ which is valid for $y<-\Im(\varphi_+(w)),$ while for $y>-\Im(\varphi_+(w))  , $ the object $R_y(z,w)=\frac{1}{z-w}$ to exponential accuracy, but this does not contribute to the integral in Eq. (\ref{actualmaineq}) . Using then Eqs. (\ref{RfromSplus},\ref{RminFinal})  one obtains: \begin{align}
&N+\log\det(\mathds{1}-K)=\\
&       =N\int_0^{-\Im\varphi_+(w)}dy\int_{}^{}\frac{dz}{2\pi\imath }\oint \frac{dw}{2\pi\imath}\frac{e^{\imath N\left[c^{-1}(\varphi_+(w)+\imath y)(z-w)\right]}-e^{\imath N(\varphi(z)
 -  \varphi_+(w)) }}{(z-w)^2}       .\nonumber\end{align}
 Performing the integral over $y$ and $z$ we obtain:
 \begin{align}
 &N+\log\det(\mathds{1}-K)=N^2\int_{\chi_\rho}\left[\frac{\varphi_+^2(w)}{2c} -\varphi_+'(w)\varphi_+(w)+\varphi_+(w)\varphi'_-(w)e^{-\imath N\varphi^J(w) }\right]\frac{dw}{2\pi},\label{MainResultGeneral}
\end{align}
where $\chi_\rho$ is the support of $\rho$. The superscript $J$ denotes the jump discontinuity of the function. Namely, for any $f$ we define $f^J(w)\equiv f(w+\imath0^+)-f(w+\imath0^+)$

 Eq. (\ref{MainResultGeneral}) represents our main result for the computation of the determinants appearing in Eq. (\ref{tocompute}). The function $\varphi$ appearing in the expression can be computed through the thermodynamic by solving the density of roots through the thermodynamic Bethe ansatz for a given value of $c$ and then substituting this in Eq. (\ref{varphiprimefull}). We believe that this result may well be extended to other Bethe ansatz models allowing the computation of Slavnov-type overlaps also there.

\section{The semiclassical limit}
To test the plausibility of our result, we take the semiclassical limit, namely $c\to0$ and compare the resulting expressions to  what is known from semiclassics \cite{belokolos:Bobenko:Algebro:Geometrical:Integrable,Bettelheim:cto0:Lieb:Liniiger,88:Flaschka:KdV:Avging,Smirnov:Quasi-Classical:qKdV,Smirnov:Babelon:Bernard:Null:Vectors}.

Let us first note an equation that will be useful later. Namely, from the definition, Eq. (\ref{varphidef}), the function  $\varphi$ has the following form at the $c\to0$ limit everywhere (except for $-c<\Im(z)<0$) and for $\Re(z)$ in the support of $\rho(x)$:
\begin{align}
&\varphi(z)=\imath\int 2\rho(z')\log\frac{z-z'}{z-z'+\imath c}dz'-z\rho_0^{-1}\label{varphiis}=\\&= 2c\int\frac{\rho(z')}{z-z'}dz'-z\rho_0^{-1}+O(c)\nonumber
\end{align}

\subsection{The Solution to the Bethe Equations at Small $c$}To obtain the small $c$ limit of the solution to the Bethe equations, first let us take the logarithm of Eq. (\ref{Bethe}):\begin{align} \imath  \rho_0^{-1}\theta_i+\int\log\frac{\theta_i-\lambda'-\imath c}{\theta_i-\lambda'+\imath c}\rho(\lambda')d\lambda'=\frac{\imath\pi}{N} (2(n_0+i)+1),\label{betheoriginal}  \end{align}
then, as usual, one takes the difference between $i+1$ and $i$ equations and multiply by \begin{align}\frac{-\imath j }{2\pi(\theta_{i+j}-\theta_i)}= \frac{-\imath N \rho(\theta_i)}{2\pi  }\label{rhostrangedef},\end{align}  an identity valid one assumes to be valid for $j$ small (we use it here mainly for $j=1$). This leads to:
\begin{align}
\frac{\rho_0^{-1}}{2\pi}+ \int\left[\frac{1}{z'-z-\imath c}-\frac{1}{z'-z+\imath c}\right]\rho(z')\frac{dz'}{2\pi\imath} =\rho(z)\label{BEtheIntegralEqforRho}
\end{align}
Let us decompose of $\rho$ into function analytic in the upper and lower half planes:
\begin{align}
\rho(z)\equiv\rho_+(z)+\rho_-(z)\label{rhodecomposition}
\end{align}
The Bethe equations (\ref{BEtheIntegralEqforRho}) then can be written as:
\begin{align} 
\frac{1}{2\pi\rho_0}+\rho_+(z+\imath c)+\rho_-(z-\imath c)=\rho_+(\lambda)+\rho_-(\lambda)\label{FullDressing}
\end{align}
Which to  leading order in $c$ reads:
\begin{align}
\rho'_+(\lambda)-\rho'_-(\lambda)= \frac{\imath\rho_0^{-1}}{2\pi c},\label{jumpofrho}
\end{align}
which is valid on the support of $\rho$. Eqs. (\ref{jumpofrho}) and (\ref{rhodecomposition}) can be solved given the condition that $\rho$ vanishes outside the support  which comprises of two intervals, $[\lambda_1,\lambda_2]\cup[\lambda_3,\lambda_4]$. This proceudure is described now. 

Consider then a function $\varphi^{(0)}(z)$ given by:
\begin{align}
\varphi^{(0)}(z)=-\rho_0^{-1}z-\imath4\pi  c\begin{cases}\rho^{(0)}_+(z) & \Im(z)>0 \\
-\rho^{(0)}_-(z) & \Im(z)<0 \\
\end{cases}.\label{varphithroughrhopm}
\end{align} 
Denote by $\mathcal{S}=[\lambda_1,\lambda_2]\cup[\lambda_3,\lambda_4]$ the support of $\rho^{(0)}$. Then we have:
\begin{align}
&\varphi'^{(0)}_+(z)+\varphi'^{(0)}_-(z)=0, \quad z\in\mathcal{S}\label{vaphi0first}\\
&\varphi'^{(0)}_+(z)-\varphi'^{(0)}_-(z)=0, \quad z\notin\mathcal{S}\\
&\varphi'^{(0)}(z)\sim-\rho_0^{-1}+\frac{C}{z^2}, \quad z\to\infty,\label{varphiatinf}
\end{align}
for some $C$ (this arises from the behavior $\rho'^{(0)}_\pm(z)\to \frac{C_\pm}{z^2}$). We may solve these conditions by:
\begin{align}
\varphi'^{(0)}(z)=-\rho_0^{-1}\frac{z^2+a_1 z+a_2}{\sqrt{R_4(z)}},
\end{align}  
where $R_4(z)=\prod_{i=1}^4(z-\lambda_i), $ and for a $a_1$, $a_2$ to be chosen below such that  Eq. (\ref{varphiatinf}) is indeed satisfied. 

An  alternative way to solve conditions (\ref{vaphi0first}-\ref{varphiatinf}) given relation (\ref{varphithroughrhopm}) is to write:
\begin{align}
\varphi^{(0)}(z)=2\imath  c\int\frac{\rho^{(0)}(x)dx}{z-x}-\rho_0^{-1} z.\label{varphiAsHilbertofrho}
\end{align}
Then one has:
\begin{align}
-2c\rho_0\int\frac{\rho^{(0)}(x)dx}{z-x}-z=\int_{\lambda_1}^z \frac{z'^2+a_1 z'+a_2}{\sqrt{R_4(z')}}dz'.
\end{align}
In order for the right hand side to be a well-defined function on the complex plane cut by the support of $\rho$, which is an obvious property of the left hand side, we must have:
\begin{align}
\int_{\lambda_1}^{\lambda_2} \frac{z^2+a_1 z+a_2}{\sqrt{R_4(z)}}dz=\int_{\lambda_3}^{\lambda_4} \frac{z^2+a_1 z+a_2}{\sqrt{R_4(z)}}dz=0,
\end{align}
or in other words, that the $b$-cycles are zero. This condition fixes $a_1$\ and $a_2$. In fact, the sum of both conditions is just the requirement that the pole at infinity of the integrand 
(considered as a differential) vanishes. This fixes $a_1$ to be equal to:
\begin{align}
a_1=-\frac{\lambda_1+\lambda_2+\lambda_3+\lambda_4}{2},
\end{align}
while the exprtession for $a_2$ involves transcendental (elliptic) functions. The density is simply related to the jump discontinuity of $\varphi^{(0)}$ due to (\ref{varphithroughrhopm}), so we have:
\begin{align}
\rho^{(0)}(z)=\int_{\lambda_{1,3}}^{z} \frac{z'^2+a_1 z'+a_2}{\sqrt{R_4(z')}}\frac{ dz'}{\imath\rho_02\pi  c}\quad   z\in[\lambda_{1,3},\lambda_{2,4}], \mbox{ respectively}.
\end{align}

The expressions above can be recast on the Jacobian by means of the Abel map. Namely, one has the map $z\to u(z)$ from the Riemann surface of the function $R_4(z)$ to the torus $\mathds{C}/\mathds{Z}\pi+\mathds{Z}\tau$. The map being given by:
\begin{align}
u(z)=\frac{1}{\omega}\int ^z_{\lambda_1} \frac{dz}{R_4(z)},\quad \omega=\frac{2}{\pi}\int^{\lambda_2}_{\lambda_1} \frac{dz}{R_4(z)},
\end{align}
and $\tau=2 \frac{u(\lambda_3)-u(\lambda_2)}{\omega}.$

Then we have:
\begin{align}
&\varphi^{(0)}(z(u))=\frac{-1}{\rho_0\omega}\int^u_0(\wp(u-u_\infty)+\wp(u+u_\infty)+\alpha )du=\\&=\frac{1}{\rho_0\omega} \left(\zeta(u+u_\infty)+\zeta(u-u_\infty)-\alpha u\right), 
\end{align}
where $\zeta$, $\wp$ are the weierstrass $\zeta$ and $\wp$ functions, respectively. Here $\alpha$ is determined by the condition that the $b$ cycles are zeros. Since the $b$ cycles correspond to $u\to u + \pi$,  and since $\zeta(u+\pi)=\zeta(u)+2\eta$, where $\eta=\zeta\left(\frac{\pi}{2} \right)$, we have $\alpha=\frac{4\eta}{\pi},$ namely:
\begin{align}
\varphi^{(0)}(z(u))=\frac{1}{\rho_0 \omega} \left(\zeta(u+u_\infty)+\zeta(u-u_\infty)-\frac{4\eta}{\pi} u\right).
\end{align}

Suppose we now remove a root, $\theta_a$,   from the support. We thus write  $\rho(x)=\rho^{(0)}(\lambda)+\frac{1}{N} \rho^{(A)}(\lambda)-\frac{1}{N}\delta(\lambda-\theta_a).$ Eq. (\ref{FullDressing}) then becomes:
\begin{align}
&\rho^{(A)}_+(\lambda+\imath c)+\rho^{(A)}_-(\lambda-\imath c)+\frac{1}{2\pi\imath}\left(\frac{1}{\lambda-\theta_a+\imath c}-\frac{1}{\lambda-\theta_a-\imath c}\right)=\rho^{(A)}(\lambda).\label{omissionEq}
\end{align} 
Note that on the right hand side there is no $\delta(\lambda-\theta_a).$ This is because $\rho$ on the right hand side of Eq. (\ref{BEtheIntegralEqforRho}) is the density of states (defined by Eq. (\ref{rhostrangedef}), in which the correction to $\rho$ due to the omission of a root does not show up since the mode number jumps now by two and the distance between the roots doubles -- which is equivalent to taking $j=2$ in Eq. (\ref{rhostrangedef})), in contrast to $\rho(z)dz$ on the left hand  side of Eq. (\ref{BEtheIntegralEqforRho}) in which is the density of occupied states wth a correction of  $-\frac{1}{N}\delta(z-\theta_a)$ due to the omission of the root $\theta_a$. 

Eq. (\ref{omissionEq}) gives then in the $c\to0$ limit:
\begin{align}
 \rho'^{(A)}_+(z)- \rho'^{(A)}_-(z)=-\frac{\imath }{ c}\delta(z-\theta_a).
\end{align}
We define similarly to what was done before:\begin{align}
\varphi^{(A)}(z)=-\imath4\pi  c\begin{cases}\rho^{(A)}_+(z) & \Im(z)>0 \\
-\rho^{(A)}_-(z) & \Im(z)<0 \\
\end{cases}.
\end{align} 
Then we have:
\begin{align}
&\varphi'^{(A)}_+(z)+\varphi'^{(A)}_-(z)=-4\pi \delta(z-\theta_a), \quad z\in\mathcal{S}\\
&\varphi'^{(A)}_+(z)-\varphi'^{(A)}_-(z)=0, \quad z\notin\mathcal{S}\\
&\varphi'^{(A)}(z)\sim\frac{C}{z^2}, \quad z\to\infty,
\end{align}

The solution to these conditions is then for $g=1$: 
\begin{align}
 \varphi^{(A)}(z)=\int^z_{\lambda_1} \frac{(z'-\theta_a)(a_0+\dots+a_{g-1}z'^{g-1})+2\imath\sqrt{\prod_{i=1}^{2g+2}(\theta_a-\lambda_i)}}{(z'-\theta_a)\sqrt{\prod_{i=1}^{2g+2}(z'-\lambda_i)}}dz'.\label{varphiAinz}
\end{align}
We
 again  demand the following condition to fix $a_0$: 
\begin{align}
\int_{\lambda_{2i-1}}^{\lambda_{2i}} \frac{(\lambda-\theta_a)a_0+2\imath\sqrt{\prod_{i=1}^{2g+2}(\theta_a-\lambda_i)}}{(\lambda-\theta_a)\sqrt{\prod_{i=1}^{2g+2}(\lambda-\lambda_i)}}=0,
\end{align}
in order to obtain a well-defined function on the upper sheet.

We now write $\varphi^{(A)}$ on the Jacobian:
\begin{align}
\varphi^{(A)}(z(u))=2\imath\int_0^{u}(\zeta(u-u_a)-\zeta(u+u_a)+\beta) du'=2\imath\log e^{\beta u } \frac{\sigma(u-u_a)}{\sigma(u+u_a)},
\end{align}
for some $\beta$ determined from the condition that  $\varphi^{(A)}(z)$ well defined on the upper sheet. Now, since $\sigma(u+\pi)=-\sigma(u)e^{2\eta(u+\frac{\pi}{2})}$ and  $\sigma(u+\tau)=-\sigma(u)e^{2\eta'(u+\frac{\tau}{2})}$  , we may  make $\varphi^{(A)}(z)$ well defined on the upper sheet by the choice of $\beta$ implied in the following:
\begin{align}
 e^{-\frac{\imath }{2}\varphi^{(A)}(z(u);u_a)}=e^{ \frac{4\eta  u_a}{\pi }u } \frac{\sigma(u-u_a)}{\sigma(u+u_a)}.\label{etothevarphiA}
\end{align}

\subsection{ The Result in the $c\to0$ Limit}

We now take:
\begin{align}
&\varphi^{(B)}(z)=- \frac{1}{z-\theta^{\rm in}_{q_2}}
\end{align}
Now a useful fact is that due to the fact tat the Hilbert transform of $\varphi$ is 0 on the cuts we have assuming the conventional method to regularize such integrals $\int \frac{f(x)}{x-y}=\dashint  \frac{f(x)}{x-y}+\imath \pi f(y)$:
\begin{align}
N\int\frac{\varphi^{(0)}(w)\varphi^{(B)}(w)}{2}\frac{dw}{2\pi}=-\frac{N}{2}\int \frac{\varphi^{(0)}(w)}{w-\theta^{\rm in}_{q_2}} \frac{dw}{2\pi }=-\frac{ N  \imath\varphi^{(0)}(\theta^{\rm in}_{q_2})}{4} 
\end{align}

We wish to compute the objects in Eq. (\ref{tocompute}). To this end we define:
\begin{align}
D\equiv\log\frac{\det( \mathds{ 1} -  K_{\bm{\theta}^{\rm out}\cup\bm \theta ^{\rm in}\setminus\theta^{\rm in}_{q_2}})}{\sqrt{ \det( \mathds{ 1} -  K_{\bm{\theta}^{\rm out}\cup\bm{\theta}^{\rm out+\varepsilon}})\det( \mathds{ 1} -  K_{\bm{\theta}^{\rm in}\cup\bm{\theta}^{\rm in+\varepsilon}}  )}}\label{Ddef}
\end{align}We can now use Eq. (\ref{MainResultGeneral}) to write:
\begin{align}
&D=\frac{N^2}{2c} \int\left[\left(\varphi^{(0)}(w)+\frac{\varphi^{(A)}(w)}{N}+\frac{c\varphi^{(B)}(w)}{N}\right)^2-\frac{1}{2}\left(\varphi^{(0)}(w)+\frac{2\varphi^{(A)}(w)}{N}\right)^2-\frac{\varphi^{(0)2}(w)}{2}\right]\frac{dw}{2\pi}\nonumber
\end{align}
Retaining only terms depending on $\theta_{q_2}^{\rm in},$  only leading terms in $c\to0$ limit and of order $N^0$, one obtains:
\begin{align}
\tilde{D}=-\frac{   \imath\varphi^{(A)}(\theta^{\rm in}_{q_2})}{2},\label{tildeD}
\end{align}
where the tilde over $\tilde{D}$ denotes that only certain terms were retained.

We now wish to compute the term $\frac{\prod_j\theta_{q_2}^{\rm in}-\theta_j^{\rm in}+\imath c}{\prod_{j\neq q_2}\theta_{q_2}^{\rm in}-\theta_j^{\rm in}}$ in Eq. (\ref{tocompute}). We shall use the following identity
\begin{align}
\frac{d}{d\lambda}\log \prod_j (\lambda-\theta_j^{\rm in})=N \psi(\lambda)\cot   N \int_{\theta_{q_2}^{\rm in}}^{\lambda } \psi(\lambda')d\lambda'\label{theidentity}
\end{align}
Where $\psi(z)$ is the analytical continuation of $\pi\rho(x)$ away from the real axis on the cuts, namely it is equal to \begin{align}\psi(z)=\pm\imath\frac{\varphi_0+\rho_0^{-1}z}{2 c},\label{psithroughphi}\end{align} which is derived from Eqs. (\ref{varphithroughrhopm}) and the fact that $\rho_\pm(x)=\frac{\rho(x)}{2}$ on the real axis. To show the equivalence of the left and right hand side, one identifies on the left hand side a analytical function in the cut plane, where on the cut one has simple poles with residue one at the point $\theta_j^{\rm in}$. On the right hand side the poles and residues match with the left hand side. Away form the real axis one has a function with asymptote $\pm N\imath\frac{\varphi(z)+\rho_0^{-1} z}{2 c }$, on the right hand side, which matches the right hand side by Eq. (\ref{varphiAsHilbertofrho}).
The base point of the integral in Eq. (\ref{theidentity}), $\theta_{q_2}^{\rm in}, $ was chosen arbitrarily for later convenience.  

Recognizing the expression in Eq. (\ref{theidentity}) as $\frac{d}{d\lambda}\log \sin \int ^\lambda N \psi $, one can write:
\begin{align}
&\frac{d}{d\lambda}\log \prod_j \frac{\lambda-\theta_j^{\rm in}+\imath c}{\lambda-\theta_j^{\rm in}}   =\frac{d}{d\lambda}\log\frac{  \sin N \int_{\theta_{q_2}^{\rm in}}^{\lambda+\imath c} \psi(\lambda')d\lambda}{  \sin N \int_{\theta_{q_2}^{\rm in}}^{\lambda} \psi(\lambda')d\lambda}, 
\end{align}
which leads to:
\begin{align}
 &\frac{\prod_j\theta_{q_2}^{\rm in}-\theta_j^{\rm in}+\imath c}{\prod_{j\neq q_2}\theta_{q_2}^{\rm in}-\theta_j^{\rm in}}=\lim_{\lambda\to\theta_{q_2}^{\rm in}} \frac{  \sin N \int_{\theta_{q_2}^{\rm in}}^{\lambda+\imath c} \psi(\lambda')d\lambda}{  \sin N \int_{\theta_{q_2}^{\rm in}}^{\lambda} \psi(\lambda')d\lambda} (\lambda-\theta_{q_2}^{\rm in})\label{Thefactor}=\\&\nonumber=\frac{e^{-\frac{N}{2c} \int_{\theta_{q_2}^{\rm in}}^{\theta^{\rm in}_{q_2}+\imath c} (\varphi_0(\lambda)+\rho_0^{-1}\lambda)d\lambda}}{2\imath \pi \rho(x)},
\end{align}
the last line being derived from the fact that $\psi(x)=\pi\rho(x)$ on the real axis and from (\ref{psithroughphi}).

Substituting  Eq. (\ref{Thefactor}, \ref{tildeD}) and (\ref{etothevarphiA}) into Eq. (\ref{tocompute}) one obtains: 
\begin{align}
\frac{\<\bm{\theta}^{\rm out}|\psi(0)|\bm \theta ^{\rm in}\>}{\sqrt{\<\bm{\theta}^{\rm out}|\bm{\theta}^{\rm out}\>\<\bm \theta ^{\rm in}|\bm \theta ^{\rm in}\>}}=C\int _{\chi_\rho} e^{ \frac{4\eta  u_a}{\pi }u(x) } \frac{\sigma(u(x)-u_a)}{\sigma(u(x)+u_a)}dx,\label{FinalSemiclassical}
\end{align}
where we have converted the sum over $\theta_{q_2}^{\rm in}$ into an integral $\int \rho(x) dx$, the density cancelling out due to Eq. (\ref{Thefactor}). Here we have a factor $C,$ which we do not compute here, relying instead on the functional form of the integral to make the comparison with known results in the semiclassical limit. 
 
\section{Comparison with Known Results from Semiclassics}
Eq. (\ref{FinalSemiclassical}) may be compared to known semiclassical results. The integrand in this equation with its typical ratio of Riemann sigma functions with a phase factor designed to make the solution periodic, can be identified with the semiclassical field $\psi(x)$ in the nonlinear Schr\"odinger 
 equation \cite{belokolos:Bobenko:Algebro:Geometrical:Integrable}. This identification can be made when one takes $u_a\to u(\infty)$, such that the difference between the arguments of the sigma functions in the numerator and denominator, respectively  is $-2u_{\infty_+}$\cite{belokolos:Bobenko:Algebro:Geometrical:Integrable,Bettelheim:cto0:Lieb:Liniiger}. Within this identification $u(x)=k X- \omega T $ is the phase of the solution, $X$\ and $T$\ being the physical time and space variables in (in contrast to $x$ which her denotes  the real axis of the auxiliary spectral surface).

It is known \cite{Smirnov:Quasi-Classical:qKdV,Smirnov:Babelon:Bernard:Null:Vectors} that  expectation values and matrix elements of in the semiclassical limit become averages over the classical solutions, with an integration measure equal to $du(x)$ rather than $dx$\ which we computed. We can attribute this discrepancy to the fact that our computation is done in the limit where $c$ is small but much larger than $1/N$. We should mention here that the limit of $c$ much smaller than $1/N$ was treated in \cite{Gorohovsky:Bettelheim:A:Derivation}, for a model closely related formally to the Lieb-Liniger model where the expected integration measure of $du(x)$ was recovered. 

We should note that in Ref. \cite{Smirnov:Quasi-Classical:qKdV}, which recovered, unlike what we have found here, the expected integration measure $du(\lambda)$, the assumption $c\ll N^{-1}$ is implicitly made. Indeed, the final result was obtained by taking the Baxter polynomial, namely the function $Q(\lambda)=\prod_i\lambda-\theta_i,$  and expanding it as:
\begin{align}
\log \frac{Q(\lambda+\imath c)}{Q(\lambda-\imath c)}       = 2\imath L \Omega_0(\lambda)+c \log \sin(L\Omega_0)+O(c^2),\label{seriesSmirnov}
\end{align} 
in a series expansion in $c$. The Baxter polynomial, $Q(\lambda)$, plays a crucial role in the approach of Ref. \cite{Smirnov:Quasi-Classical:qKdV} since it serves as the wave function in separated coordinates, and as such expectation values are represented as averages with square modulus of the wave-function as integration measure. For the series above, Eq. (\ref{seriesSmirnov}), to converge, one must assume $cL\ll 1,$ which is equivalent to $cN\ll1$ given that $L$ is of order $N$. For example, the second term in the expansion is not small, since for $\Omega_0$ that has an imaginary part  of order $1$, the function $\log \sin(L\Omega_0)$ can be approximated, to exponential accuracy, as  $\pm \imath L\Omega_0-\log 2\imath,$ where the sign depends on whether the imaginary part of $\Omega_0$ is positive or negative, respectively. Thus, to make $c \log \sin(L\Omega_0)$ small one must have $cL\ll1.$  In fact it is the term
$c \log \sin(L\Omega_0)$ in the expansion above that is responsible for the term $\frac{1}{\sqrt{R_4(\lambda)}}$ that turns $d\lambda$ into $du(\lambda)$ in the final result.

To show that recovering the semiclassical measure of integration in the case where $c$ is allowed to become small than $1/N$ is plausible, we consider here what roughly changes in such a  limit. It seems that the crucial factor that changes in this limit is the fact that the density of the Bethe roots goes to zero at the edges of the support of the density. This is to be contrasted with the case for finite $c$ in which the density of Bethe roots has a jump discontinuity at the edge of the support. Indeed for finite $c$, Eq. (\ref{FullDressing}) may be solved for $z$ real and less than zero (we assume the support is from $-\infty$ to $0$) and $z\ll c$  by letting $\rho_\pm(z)= \pm \frac{\imath}{2\pi^2\rho_0}\log(z).$

Our method for computing the determinants appearing in the overlaps, Eq. (\ref{tocompute}), is actually inconsistent with the density going to zero at the edge of the support. This is because this is accompanied by $\varphi$ going to zero at the edge of the support and our method relied on the steepest descent method being applied to integrands and kernels of the form $e^{\imath N\varphi}$. In fact, the computation of $R(z,w)$ relies on $e^{\imath N \varphi(w)}$ being exponentially large. In order to deal with small $c, $  for $w$ near the edge of the support (in which case $e^{\imath N \varphi(w)}$ is of order $1$), one may apply the method of Ref. \cite{Kostov:Bettelheim:SemiClassical:XXX}.

Within the approach of Ref. \cite{Kostov:Bettelheim:SemiClassical:XXX}, the leading order contribution to the logarithm of the relevant determinants is given by:  
\begin{align}
\log \det \mathds{1}-K=-\frac{1}{c}\oint{\rm Li}_2\left(1-e^{N\imath \varphi (w)}\right) \frac{dw}{2\pi\imath}.
\end{align}
Expanding for $\varphi(w)$ small one obtains:
\begin{align}
\log \det \mathds{1}-K=\oint \log\left( N\imath \varphi (w)\right)\frac{N\varphi(w)dw}{2\pi\imath c}=\int   \log\left( N\imath \varphi (w)\right)dn,
\end{align}
where $n$ is the mode number of the Bethe roots. Considering now the contribution to $\log D$ from such terms one encounters a terms of the form:
\begin{align}
\log D\sim\int   \log\left( N\imath \varphi^{(B)} (w)\right)dn\sim A \sum \log(z-\lambda_i). 
 \end{align}
 This equation is not exact, but only means to point out that $\log D$ will have logarithmic divergencies at the edges of the support of the Bethe roots, which we denote by $\lambda_i$. The prefactor, $A$, in front of these logarthmic singularities are to be determined by a much more careful calculation. Neverthless, the point here is that once these singularities are cut off at finite $c$ they may disappear from the final result as we saw in this paper where $c$ was considered small but does not scale with $N$, such that with respect to $N$ it may be considered finite. We speculate that this is the case with the factor $\frac{1}{\sqrt{R_4(z)}}$ which does not appear in  the limit $1/N\ll c \ll 1$, namely in essence that $A$\ turns out to be $-1/2$  in the $c\sim 1/N$ limit.

\section{Conclusion}

We have presented a method to compute matrix elements in the Lieb-Liniger model. This model is one of the simplest examples of the Bethe ansatz at work, with the absence of strings, and the simple algebraic R-matrix. Possible directions to generalize our result would be to consider one of the many models for which the Slavnov matrix representation exists for its overlaps in the form suggested in Ref. \cite{Kostov:Inner:Product:Domain:Wall}, which is the starting point of this paper. Indeed, the method uses here allows to compute the overlaps for a wide range of scenarios.  One can often convert the computation matrix elements of operators into the computation of Slavnov-type overlaps as was done here following  Refrs. \cite{Calabrese:Lieb:Liniger:Form:Factors,Gorohovsky:Bettelheim:Coherence:Announce,Gorohovsky:Bettelheim:A:Derivation}, but a similar procedure may be possible in additional models, such that the method here may be applied to compute interesting physical observables. After all, physical observables in and out of equilibrium can invariably be written through matrix elements and expectation values. 

The current paper thus can be viewed as part of a wider program to further understand integrable systems, beyond equilibrium properties, by using the large $N$ limit in the computation of physical observables, by mapping the latter into the computation of {\it\ functional} determinants (as was done here by defining the object $R(z,w)$, the resolvent in function space), and finding the resolvent by a steepest descent method.

\section{Acknowledgement}
I\ would like to acknowledge money from  ISF\ grant number 
 1466/15.

\end{document}